\documentclass[nofootinbib,preprint,superscriptaddress]{revtex4}

\usepackage{graphicx}  %
\usepackage{amsmath}
\usepackage{bm}  %
\usepackage{ulem}  %
\usepackage{verbatim} %
\usepackage{color} %
\usepackage{hyperref}
\usepackage{multirow}
\allowdisplaybreaks

\newcommand{\be}{\begin{equation}}
\newcommand{\ee}{\end{equation}}

\def\lag{\mathcal{L}}

\def\mt{{\mathcal{T}}}
\def\vep{\varepsilon}
\def\<{\langle}
\def\>{\rangle}

\newcommand{\bea}{\begin{eqnarray}}
\newcommand{\eea}{\end{eqnarray}}
\newcommand{\beq}{\begin{equation}}
\newcommand{\eeq}{\end{equation}}
\newcommand{\bqa}{\begin{eqnarray}}
\newcommand{\eqa}{\end{eqnarray}}

\def\mqo2{{\!\!\!}}
\def\SU{{\text{SU}}}
\def\tr{\mathrm{tr}}

\def\SO{{\text{SO}}}

\def\hatp{{\hat{p}}}
\def\hatP{{\hat{P}}}
\def\hatm{{\hat{M}}}

\usepackage{relsize}
\def\babar{\mbox{\slshape B\kern-0.1em{\smaller A}\kern-0.1em
    B\kern-0.1em{\smaller A\kern-0.2em R}}}

\begin{document}

\vspace*{1cm}
\title{Operator Counting and Soft Blocks in  Chiral Perturbation Theory}
\def\DukeU{Department of Physics, Duke University, Durham, NC 27705, USA}
\def\Argonne{\mbox{High Energy Physics Division, Argonne National Laboratory, Lemont, IL 60439, USA}}
\def\NU{\mbox{Department of Physics and Astronomy, Northwestern University, Evanston, IL 60208, USA}}

\author{\vspace{.5cm}Lin Dai}
\email{lin.dai@pitt.edu}
\affiliation{\DukeU}
\author{Ian Low}
\email{ilow@northwestern.edu}
\affiliation{\Argonne}
\affiliation{\NU}
\author{Thomas Mehen}
\email{mehen@phy.duke.edu}
\affiliation{\DukeU}
\author{Abhishek Mohapatra}
\email{abhishek.mohapatra@duke.edu}
\affiliation{\DukeU}
\date{\today}

\begin{abstract}
\vspace*{.5cm} 
Chiral perturbation theory (ChPT) is a low-energy effective field theory of QCD and also a nonlinear sigma model based on the symmetry breaking pattern ${\rm SU}(N_f)\times {\rm SU}(N_f)\to {\rm SU}(N_f)$. In the limit of massless $N_f$ quarks, we enumerate the independent operators without external sources in ChPT using an on-shell method, by counting and presenting the  soft blocks at each order in the derivative expansion, up to ${\cal O}(p^{10})$. Given the massless on-shell condition and total momentum conservation, soft blocks are homogeneous polynomials of kinematic invariants exhibiting the Adler's zero when any external momentum becomes soft and vanishing.  In addition, soft blocks are seeds for recursively generating all tree amplitudes of Nambu-Goldstone bosons without recourse to ChPT, and in one-to-one correspondence with the "low energy constants" which are the Wilson coefficients.  Relations among operators, such as those arising from equations of motion, integration-by-parts, hermiticity, and symmetry structure, manifest themselves in the soft blocks in simple ways. We find agreements with the existing results up to NNNLO, and make a prediction at N$^4$LO.
\end{abstract}
\smallskip
\maketitle

\tableofcontents

\section{Introduction}\label{sec:Intro}

Chiral perturbation theory describes the low-energy dynamics of QCD \cite{Weinberg:1968de}, in particular interactions of mesons and baryons below the energy scale $\Lambda\sim {\cal O}(1\ {\rm GeV})$, and is an integral component of our understanding of many nuclear processes.
Conceptually ChPT is the earliest, and perhaps also the most elegant, example employing the modern technique of effective field theory (EFT) \cite{Weinberg:1978kz}, which later found wide-ranging applications in diverse topics.

The success of ChPT lies in the key observation that, in most cases, symmetry alone is sufficient to capture the long wavelength dynamics of a physical system; short wavelength fluctuations are encoded in an infinite number of ``coupling constants,'' or Wilson coefficients in the language of renormalization group evolution, which are given as inputs in the EFT. In order to achieve any predictive power, a power counting scheme must be supplied to organize the relative importance of different coupling constants so that at low energies only a small number of them are needed to describe experimental processes relevant at a certain energy scale and a given precision.

In ChPT, the relevant symmetry is the chiral symmetry of light quarks, $\SU(N_f)_L\times SU(N_f)_R$, where $N_f$ is the number of light quark flavors and can be 2 or 3, depending on if the strange quark is included.  In our world chiral symmetry is broken spontaneously to the diagonal group $\SU(N_f)_V$ due to confinement. The degrees of freedom in ChPT are not quarks and gluons, but are mesons (such as pions), nucleons and the photon. In particular, pions play a crucial role in ChPT because they are the Nambu-Goldstone bosons for the spontaneously broken chiral symmetry.\footnote{This is the case for $N_f=2$. For $N_f=3$,  kaons and $\eta$ mesons are also included.} 

A common power counting in ChPT is  the momentum expansion -- operators in the effective Lagrangian are organized in increasing powers of momenta, i.e. derivatives in position space, and the coupling constants are called low-energy constants (LEC's). For an experimental process carrying the typical momentum $p$, the effect of an effective operator carrying $n$ derivatives is characterized by $(p/\Lambda)^n$, where $\Lambda$ is a mass scale dictating the convergence of the momentum expansion. So for small momentum transfer, $p\ll \Lambda$, only a few effective operators need to be included; those carrying a high number of derivatives can be neglected.  

In addition to $\Lambda$, there is a second  mass scale in ChPT, 
\be
\langle 0| J^\mu_A (x) | \pi \rangle \propto  e^{ip\cdot x} \ f_\pi \, p^\mu\ ,
\ee
where $J^\mu_A(x)$ is the axial current for $\SU(N_f)_A$, $p^\mu$ is the pion momentum and $f_\pi$ is the pion decay constant. This is the famous PCAC (Partially Conserved Axial Current) hypothesis in low-energy QCD \cite{Dashen:1969ez}, which  states that the axial current would be exactly conserved if the pion were massless. At a fixed order $n$ in the momentum expansion, there are actually an infinite number of operators carrying $n$ derivatives and an arbitrary even power of pion fields $(\pi/f_\pi)^{2k}$. However this doesn't imply there are an infinite number of LEC's at a fixed derivative order, because operators carrying  different powers of $\pi/f_\pi$ are often related to one another by the broken symmetry group in a nonlinear way. One can choose a parameterization for the nonlinearly realized broken symmetry and resum the corresponding $1/f_\pi$ expansion into a compact analytic form, thereby arriving at an operator that is invariant under the complete $\SU(N_f)_L\times \SU(N_f)_R$. Each invariant operator is multiplied by a LEC, which needs to be measured experimentally. In the end, the number of LEC's is finite at each order in the momentum expansion and independent of the nonlinear parameterization. 

At the leading ${\cal O}(p^2)$, the lowest order term  in the $1/f_\pi$ expansion is the kinetic term of the pions, and all higher order terms in $1/f_\pi$ are uniquely related to the pion kinetic terms by the broken symmetry, leading to one invariant operator carrying two derivatives. There is no LEC at this order because the coefficient of the ${\cal O}(p^2)$ operator is fixed by canonically normalizing the scalar kinetic term. One popular nonlinear parameterization is \cite{Weinberg:1978kz}
\be
{\cal L}^{(2)} = \frac{1}{2}\ \frac{\partial_\mu \pi^a \partial^\mu \pi^a}{(1+\pi^a \pi^a/f_\pi^2)^2} \ , \quad a=1,2,3\ .
\ee
Since the pions transform as the adjoint of $\SU(N_f)_V$, the invariance under the unbroken group is explicit. It is much less obvious that the above equation is also invariant under the broken $\SU(N_f)_A$. At ${\cal O}(p^4)$ there are four invariant operators in general and, therefore, four LEC's. At higher orders in the momentum expansion, it is customary to rescale each derivative $\partial_\mu \to \partial_\mu/\Lambda$ and each pion field $\pi^a\to \pi^a/f_\pi$, then the structure of ChPT reads
\be
{\cal L}_{\rm ChPT} =\Lambda^2 f^2 \left[\widetilde{\cal L}^{(2)}(\partial/\Lambda, \pi/f_\pi)+\widetilde{\cal L}^{(4)}(\partial/\Lambda, \pi/f_\pi) + \cdots \right]\ , \label{eq:chptscale}
\ee
where $\widetilde{\cal L}^{(n)}$ contains only operators carrying $n$ derivatives. Moreover, demanding that quantum corrections to Eq.~(\ref{eq:chptscale}) cannot be larger than the natural size of each operator gives rise to the relation,
\be
\Lambda \sim 4\pi f_\pi\ .
\ee
This is the expectation from the so-called Naive Dimensional Analysis \cite{Manohar:1983md}. 

The construction of a minimal basis for invariant operators in EFT's is obviously an important question and notoriously tricky to deal with, due to the fact that different operators can sometimes lead to identical S-matrix elements. Two well-known examples are operators related by integration-by-parts and equation-of-motion. For the Standard Model (SM), there is a vast amount of literature in pursuing all higher dimensional operators consistent with (approximate) symmetries of the SM; some influential works can be found in Refs.~\cite{Weinberg:1980bf,Buchmuller:1985jz,Grzadkowski:2010es}. The effort intensified in recent years, mainly driven by the need to confront the SM with the ever improving experimental precision. This line of investigation has also led to interesting theoretical questions and new underlying structures of EFT's \cite{Alonso:2014rga,Lehman:2015via,Cheung:2015aba,Henning:2015daa,Henning:2017fpj,Kobach:2018nmt}. For ChPT, there are additional complications due to operator relations imposed by the nonlinearly realized broken symmetry, making the counting of operators more involved. In spite of the difficulty,
the construction of the effective Lagrangian up to ${\cal O}(p^6)$ (NNLO) was completed a while ago \cite{Gasser:1983yg,Fearing:1994ga,Bijnens:1999sh} and ${\cal O}(p^8)$ (NNNLO) Lagrangian was obtained  recently \cite{Bijnens:2018lez}.

In this work we would like to take up a relatively modest task: counting the number of independent parity-even operators in ChPT, in the massless quark limit and external sources turned off, up to ${\cal O}(p^{10})$ (N$^4$LO) and provide an ``on-shel'' basis for the operators.  This is less ambitious than building the complete N$^4$LO effective Lagrangian, and our study could serve as a starting point toward that goal. Moreover our methodology is new, in the sense that we will employ the ``soft bootstrap'' approach developed in the modern S-matrix theory \cite{Cheung:2015ota,Elvang:2018dco,Low:2019ynd}, which aims to (re)construct effective field theories from the peculiar soft limit of  S-matrix elements. The approach relies on finding higher point (pt) tree amplitudes from lower pt amplitudes by imposing unitarity and soft limit,\footnote{The effort was actually initiated long time ago using the current algebra approach \cite{Susskind:1970gf,Ellis:1970nn}, which is quite cumbersome and difficult to systematize.} which can be implemented easily via the soft recursion relation \cite{Cheung:2015ota}. The seed amplitude for the recursion relation is called the soft block and in one-to-one correspondence with the invariant operator at each order in the momentum expansion \cite{Low:2019ynd}. In practice, the soft blocks are the lowest pt contact interaction from each invariant operator when all external legs are on-shell. Working with soft blocks has the advantage that the redundancies in operator relations manifest themselves in the on-shell method in simple ways. 

This work is organized as follows. In Section \ref{sec:NLSM} we provide a brief overview of recent progress in understanding the interactions of Nambu-Goldstone bosons from the soft limit of on-shell amplitudes, after which we discuss  in Section \ref{sect:softblocks}  properties of soft blocks that are relevant for our computation. A basis for the kinematic invariants we employ is also given there. Our main results are contained in Section \ref{sec:comuting}, where soft blocks up to ${\cal O}(p^{10})$ are presented. We also include an ancillary {\tt Mathematica} notebook detailing the calculation. Then we summarize in Section \ref{sect:summary}.
Several long expressions, as well as an expository description of the {\tt Mathematica} notebook, are relegated to the Appendices.

\section{A brief overview of IR formulation}\label{sec:NLSM}

The construction of higher order effective operators in ChPT is usually based on the general formalism developed by Callan, Coleman, Wess, and Zumino (CCWZ) in Refs.~\cite{Coleman:1969sm,Callan:1969sn}, which is a top-down approach in that it requires specifying a broken group $G$ in the ultraviolet (UV) and an unbroken group $H$ in the infrared (IR). In the context of ChPT, the approach has been reviewed extensively; see for examples Refs.~\cite{Leutwyler:1993iq,Ecker:1994gg,Pich:1995bw,Scherer:2002tk,Bernard:2006gx}. Here we give a brief overview of recent progress in describing interactions of Nambu-Goldstone bosons from the IR, without recourse to CCWZ, by specifying the soft limit of S-matrix elements. 

The starting point of the IR formulation of ChPT is the Adler's zero condition \cite{Adler:1964um}, which states that the on-shell amplitudes of Nambu-Goldstone bosons must vanish when one external momentum $p$ is taken to be soft: $p\to \tau p$ and $\tau\to 0$. Although usually phrased as an on-shell statement and initially proven using the old-fashioned current algebra technique, the Adler's zero condition actually follows from the quantum Ward identity of a shift symmetry  on the Nambu-Goldstone field $\pi^a$ in the Lagrangian formulation \cite{Low:2015ogb},
\be
\pi^a \to \pi^a +\varepsilon^a +\cdots \ ,
\ee
where $\varepsilon^a$ is an infinitesimal constant shift. Terms neglected above arise from the non-trivial vacuum structure of the unbroken group $H$ \cite{Low:2018acv} and vanish trivially for unbroken U(1) (sub)group.

More specifically, let's consider a set of scalars $\{\pi^a\}$ transforming as a linear representation of an unbroken group $H$ with generators $T^r$. It is always possible to choose a basis such that $T^r$ is purely imaginary and anti-symmetric \cite{Weinberg:1996kr}:
\be
[T^r]_{ab} = -[T^r]_{ba} \ , \qquad (T^r)^* = - T^r \ .
\ee
The shift symmetry enforcing the Adler's zero condition under the constraints of the $H$-symmetry is highly nonlinear \cite{Low:2014nga,Low:2014oga},
\be
\label{eq:shiftall}
\pi^a \to \pi^a + \left[\sqrt{\mt} \cot \sqrt{\mt}\right]_{ab}  \vep^b\ , \qquad
\left[\mt\right]_{ a  b} = \frac{1}{f^2} [T^r]_{ a c} [T^r]_{d b}\  \pi^c \pi^{d} \ ,
\ee
where $f$ is the "decay constant" of the Nambu-Goldstone boson. The effective Lagrangian invariant under Eq.~(\ref{eq:shiftall}) can be constructed using the following two objects:
\be
d_\mu^a =  \frac{1}{f} \left[F_1 (\mt)\right]_{ab} \partial_\mu \pi^b \ , \qquad
E_\mu^r =  \frac{1}{f^2}  \partial_\mu \pi^a  \left[F_2 (\mt)\right]_{ab} \left[T^r\right]_{bc} \pi^c\ ,\label{eq:irdefe}
\ee
where
\be
F_1 (\mt) = \frac{\sin \sqrt{\mt}}{\sqrt{ \mt} } \ , \qquad F_2 (\mt ) = -\frac{2i}{\mt} \sin^2 \frac{\sqrt{\mt}}{2} \  .
\ee
In the above $d_\mu^a$ transforms covariantly under the shift symmetry in Eq.~(\ref{eq:shiftall}),  while $E_\mu^i$ transforms in the adjoint representation of $H$ like a ``gauge field,''
\bea
\label{eq:dmutransf}
d^a_\mu  &\to& \left[h_A (\vep, \pi)\right]_{ab}\  d^b_\mu\ ,\\
E_\mu^r T^r &\to&\  h_A (\vep, \pi)\  E_\mu^r T^r\ h_A^\dagger (\vep, \pi) - i\, h_A (\vep, \pi)\ \partial_\mu h_A^\dagger (\vep, \pi)\ ,\label{eq:eitransf}
\eea
where the specific form of $h_A (\vep, \pi)$ does not concern us here. The important observation here is that these two building blocks are constructed using only IR data, without reference to the broken group $G$, as long as the Closure condition is satisfied \cite{Low:2014nga},
\be
\left[T^r \right]_{ab}\left[T^r \right]_{cd}+\left[T^r \right]_{ac}\left[T^r \right]_{db}+\left[T^r \right]_{ad}\left[T^r \right]_{bc} = 0\ ,\label{eqclcd}
\ee
which applies to many commonly encountered situations, such as the adjoint of SU$(N)$ group or the fundamental of SO$(N)$. Another important comment is the normalization of the decay constant $f$ is undetermined in the IR formulation, in that the effective Lagrangian constructed this way is invariant under the rescaling of $f\to c f$ for arbitrary $c$. Moreover, a purely imaginary $c$ corresponds to a non-compact coset, while a real $c$ arises from a compact coset \cite{Low:2014nga}.

In ChPT, $H={\rm SU}(N_f)$ and $\pi^a$'s transform under the adjoint representation.  We can then write $d_\mu = d_{\mu}^a T^a$ and the leading two-derivative operator is unique:
\bea
\lag^{(2)} = \frac{f^2}{2}  {\rm tr} \left(d_\mu d^{\mu}\right) \ , \label{eq:nlsmir}
\eea
where the coefficient is fixed by canonical normalization of the scalar kinetic term.  The leading order (LO) equation of motion is
\be
\nabla_\mu d^\mu = 0 \ ,
\ee
where $\nabla_\mu d_\nu \equiv \partial_\mu d_\nu + i [E_\mu, d_\nu]$. At the next-to-leading order (NLO), there are four independent operators carrying four derivatives,
\bea
O_1 &=& \left[ {\rm tr} (d_\mu d^\mu)\right]^2\ ,\label{eq:o4nt1}\\
O_2 &=& \left[ {\rm tr} (d_\mu d_\nu)\right]^2 \ ,\label{eq:o4nt2}\\
O_3 &=& \tr ( [ d_\mu, d_\nu]^2 )\ ,\\
O_4 &=& \tr ( \{ d_\mu, d_\nu \}^2 )\ ,
\eea
and the NLO effective Lagrangian is
\be
\lag^{(4)} = \sum_{i=1}^4 L_{4,i} \, O_i \ . \label{eq:lecdef}
\ee
In the above the coefficients $L_{4,i}$ are the low-energy constants (LEC's) at NLO. In principle one could write several more operators at this order, but they are all equivalent to the above list upon integration-by-parts, equations of motion at LO, and the following two important identities,
\bea
\nabla_{[\mu} d_{\nu] } = 0\ , \qquad E_{\mu \nu} &\equiv& -i [ \nabla_\mu, \nabla_\nu]=  -i[d_{\mu}, d_{\nu}]\ , \label{eq:covd0}
\eea
which can also be derived in the IR \cite{Low:2019ynd}.\footnote{In the coset construction  the identities follow from the Maurer-Cartan equation \cite{DHoker:1995mfi}.} The list of operators can be further reduced if we specialize to $N_f=2, 3$.

At NNLO, operators in ChPT for a general $N_f$ are first presented in Ref.~\cite{Bijnens:1999sh} and later clarified in Ref.~\cite{Bijnens:2018lez}, while the N$^3$LO Lagrangian was studied in Ref.~\cite{Bijnens:2018lez}. In this work we shall focus on ChPT in the limit of massless $N_f$ quarks and without external sources.

\begin{figure}[t]
\begin{center}
 \includegraphics[width=14cm]{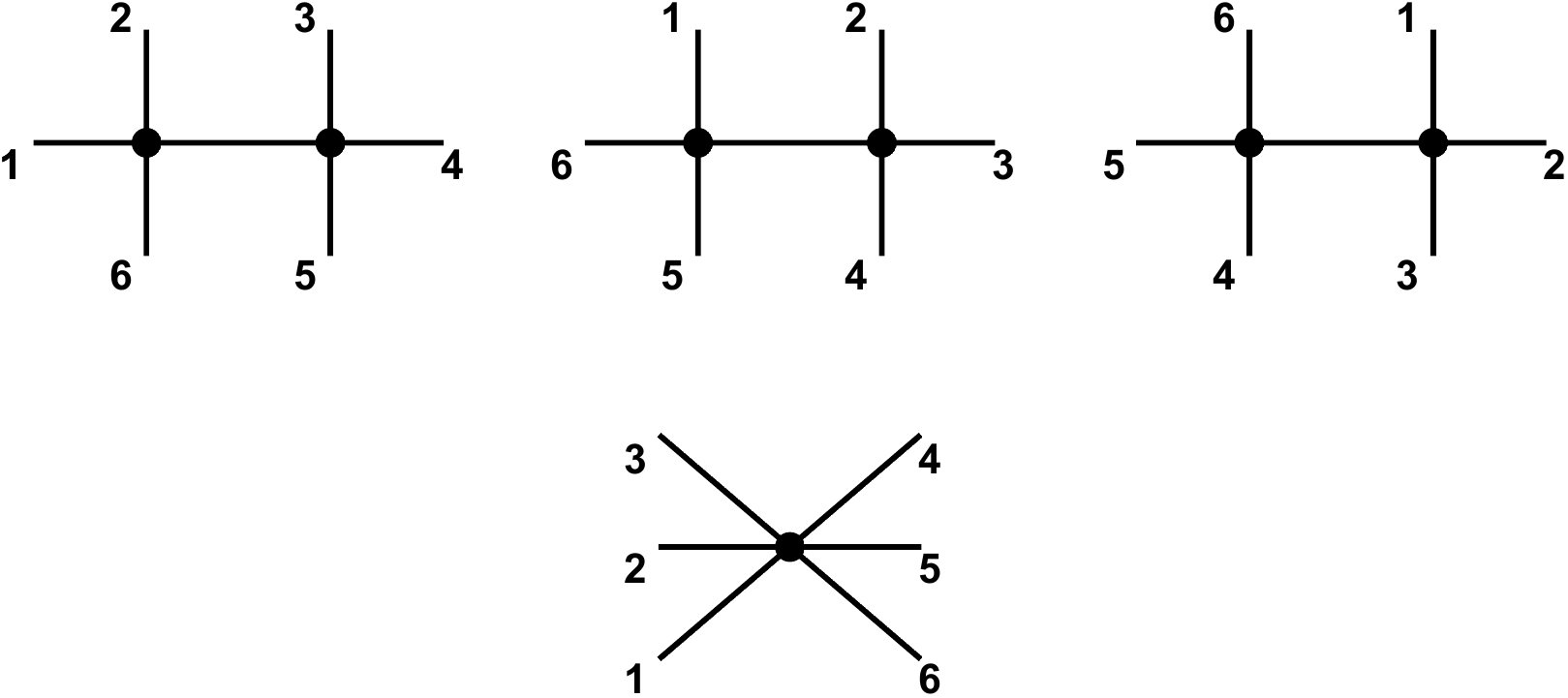}
\end{center}
\caption{The Feynman diagrams for the $6$-pt pionic amplitudes. The $6$-pt contact interaction in the bottom row is required to satisfy the  Adler's zero condition.}
\label{fig:6pt}
\end{figure}

The preceding IR construction in the Lagrangian approach has an on-shell counterpart, which is pioneered by Susskind and Frye in Ref.~\cite{Susskind:1970gf}. Starting from the 4-point (pt) amplitudes of pions, Susskind and Frye constructed the 6-pt pionic amplitudes, by introducing a 6-pt contact interaction so that the Adler's zero condition is satisfied; see Fig.~\ref{fig:6pt}. At the time up to 8-pt amplitudes were constructed this way, and the relation of this approach to the then popular current algebra was clarified in Ref.~\cite{Ellis:1970nn}

The modern approach, which is developed in  Refs.~\cite{Cheung:2015ota,Cheung:2016drk} and  goes by the name of ``soft bootstrap,'' relies on the powerful techniques of on-shell recursion relations for scalar theories with vanishing single soft limit. While ``normal'' recursion relations for Yang-Mills theories and gravity rely on the amplitudes vanishing at infinity when deforming the real momentum into the complex plane, on-shell amplitudes for scalar theories do not have such nice properties. Instead, ``soft'' recursion relations make use of the vanishing single soft limit of the amplitudes to construct an integrand that vanishes at complex infinity. In essence, the recursion relation works by searching for an expression, constructed from lower-pt amplitudes, which possesses the correct single soft limit and factorization property of a higher-pt amplitudes.

Focusing on ChPT, where the Nambu-Goldstone bosons transform as the adjoint of the unbroken $\SU(N_f)$, it will be convenient to introduce flavor-ordered partial amplitudes \cite{Kampf:2013vha}, which are akin to the color-ordered partial amplitudes in the $\SU (N)$ Yang-Mills theory. The full amplitude at the leading ${\cal O}(p^2)$ can be written as
\bea
M^{a_1 \cdots a_{n} } (p_1, \cdots, p_{n}) \equiv \sum_{\sigma \in S_{n-1}} \mathcal{C}^{ a_{\sigma (1)} \cdots a_{\sigma (n-1)} a_n} M( \sigma(1), \cdots, \sigma (n-1), n )\ ,\label{eq:stfod}
\eea
where 
\bea
\mathcal{C}^{a_1  a_2 \cdots a_n } = \tr \left( T^{a_1} T^{a_2} \cdots T^{ a_n} \right)\ , \label{eq:ffat}
\eea  
 is the flavor factor, $\sigma$ is a permutation of indices $\{1, 2, \cdots, n-1\}$ and $T^a$ is the generator of $\SU (N_f)$ group.  The full amplitude has permutation invariance among all external legs, while the partial amplitude $M(1,2,\cdots n)$ is  invariant only under {\it cyclic} permutations of the external legs due to  Eq.~(\ref{eq:ffat}). Beyond LO, the flavor factor $\mathcal{C}^{a_1  a_2 \cdots a_n }$ may involve multi-trace structure and the  partial amplitudes will only have the corresponding partial cyclic invariance.

At a fixed order in the derivative expansion, the soft bootstrap program constructs higher-pt amplitudes from lower-pt amplitudes through  the soft recursion relation \cite{Cheung:2015ota}, which utilizes an all-leg shift for the external momenta of  $M_n$,
\bea
\label{eq:alllegshift}
p_i \to \hatp_i = (1-a_i z)\, p_i\ ,
\eea
where $z$ is a complex shift parameter and total momentum conservation requires,
\bea
\sum_{i=1}^n a_i\, p_i^{\mu} = 0\label{eq:smcc} \ ,
\eea
where we have adopted the convention that all momenta are incoming.
The deformed amplitude $\hatm_n (z)$ with momentum variables $\hatp_i$ is still an on-shell amplitude, in the sense that all external momenta remain on-shell, and taking the soft limit of $p_i$ corresponds to setting $z \to 1/a_i$. In $D$-dimensional spacetime, non-trivial solutions for $a_i$ exist when $n-1 \ge D+1$, and the number of distinct solutions is $n-D-1$ \cite{Cheung:2015ota}.

\begin{figure}[t]
\begin{center}
 \includegraphics[width=14cm]{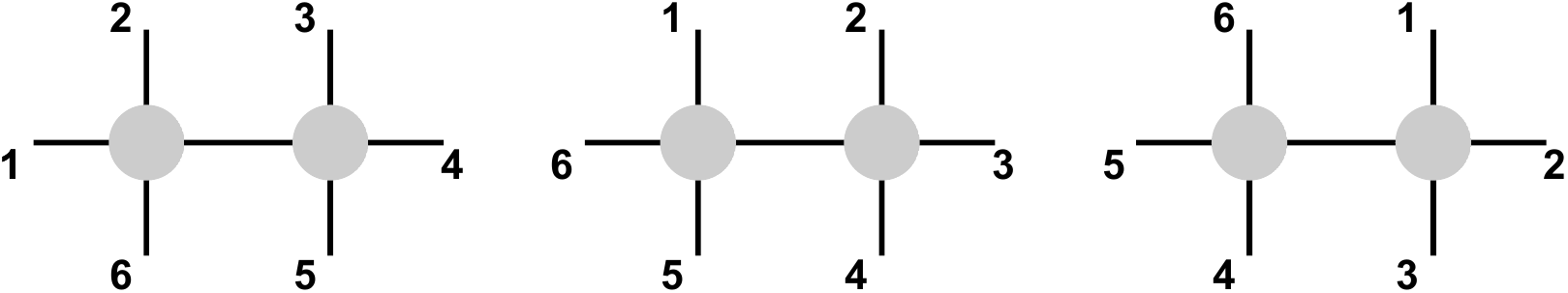}
\end{center}
\caption{ The three factorized channels for the $6$-pt single trace amplitude $M\left(1,2,3,4,5,6\right)$. The contribution of these three diagrams are equivalent to those from the four Feynman diagrams shown in Fig.~\ref{fig:6pt}.}
\label{fig:st6pfc}
\end{figure}

The deformed amplitudes $\hatm_n (z)$ does not vanish at the complex infinity $z\to \infty$. However, given that  the Nambu-Goldstone amplitude in ChPT  must satisfy the Adler's zero condition, $\hatm_n(z)\sim \hat{p}$ for $\hat{p}\to 0$  \cite{Adler:1964um}, it is possible to define a soft factor  \cite{Cheung:2015ota},
\bea
F_n(z) \equiv \prod_{i=1}^n (1-a_i z)\ ,\label{eq:sffn}
\eea
so that the Cauchy integral,
\bea
\oint \frac{dz}{z} \frac{\hatm_n (z) }{F_n(z)}=0\ . \label{eq:cauc}
\eea
contains poles only $z = 0$ and the factorization channel $I$, where the shifted momentum of an internal propagator $\hatP_I(z)$ becomes on-shell:
\be
\hatP_I(z) = \sum_{i \in I} p_i -z\, \sum_{i\in I} a_i \, p_i \ , \qquad \qquad  \hatP_I(z)^2=0 \ .
\ee
Cauchy's theorem then relates the $n$-pt amplitude $M_n$, which is the residue at $z=0$, to the other residues at $\hatP_I^2(z_I^\pm)=0$:
\bea
M_n = \hatm_n (0) = -\sum_{I, \pm} \frac1{P_I^2} \frac{\hatm_L^{(I)}  (z_I^\pm) \hatm_R^{(I)}  (z_I^\pm) }{F_n(z_I^\pm) (1-z_I^\pm/z_I^\mp )}\ , \label{eq:sbrr}
\eea
where $M_L$ and $M_R$ are the two lower-pt sub-amplitudes associated with the factorization channel $I$. As an example, the three factorization channels of 6-pt partial amplitudes are shown in Fig.~\ref{fig:st6pfc}, which is equivalent to the contribution of the four diagrams in Fig.~\ref{fig:6pt}.

Eq.~(\ref{eq:sbrr}) is the foundation of the soft bootstrap program, which determines higher-pt amplitudes from lower-pt amplitudes at a fixed order in the derivative expansion, by imposing the desired soft limit on the amplitudes. The nonlinear shift symmetry in Eq.~(\ref{eq:shiftall}) is the algebraic realization of soft bootstrap in the Lagrangian formulation, since the Adler's zero condition follows from the Ward identity corresponding to Eq.~(\ref{eq:shiftall}) \cite{Low:2017mlh}.

\section{Power Counting and Soft Blocks }
\label{sect:softblocks}

Since soft bootstrap constructs higher-pt amplitudes from lower-pt amplitudes, an input is needed to serve as the seed for soft bootstrap. These inputs are the soft blocks. A soft block ${\cal S}^{(k)}(p_1,\cdots, p_n)$ is a contact interaction carrying $n$ scalars and $k$ derivatives that satisfies the Adler's zero condition when all external legs are on-shell. In other words, each soft block is a homogeneous polynomial of kinematic invariants $\{s_{ij}\equiv 2 p_i\cdot p_j;\ i,j=1,\cdots, n\}$ of degree $k/2$,  which satisfy the Adler's zero condition. Note that for scalar field theories $k$ is always even.

For each derivative order $k$, the number of soft blocks is finite. First of all, no non-trivial soft blocks exist for $n\le 3$, because of total momentum conservation. For $n>3$, let's perform the all-leg-shift in Eq.~(\ref{eq:alllegshift}) and the soft block now becomes a polynomial of degree $k$ in $z$: $\hat{\cal S}^{(k)}=\hat{\cal S}^{(k)}(z)$. Taking the single soft limit in momentum $p_i$ is now equivalent to taking $z\to 1/a_i$. The Adler's zero condition then implies  
\be
\hat{\cal S}^{(k)}(1/a_i)=0, \qquad i=1, \cdots, n \ .
\ee
That is $1/a_i$ is a root of $\hat{\cal S}^{(k)}(z)$. However, in $D=4$  there exists non-trivial solutions for $a_i$'s only when the number of external legs $n \ge 6$ \cite{Cheung:2015ota}. So we will distinguish two separate cases: i) $k\le 4$ and ii) $k\ge 6$. 

In case i), $\hat{\cal S}^{(k)}(z)$ is a polynomial in $z$ of degree $k\le 4$, which has at most 4 distinct roots. Therefore soft blocks with $n\ge 6$ cannot exist. For $n=4,5$, the $a_i$'s are trivial and the argument fails. So soft blocks with 4 or 5 external legs could exist in principle and are indeed found in Ref.~\cite{Low:2019ynd}. 

In case ii), $\hat{\cal S}^{(k)}(z)$ is a polynomial in $z$ of degree $k\ge 6$, so soft blocks with $n\le k$ could be there. The goal of this study is to enumerate all soft blocks for $k\le 10$.

In the Lagrangian approach, the soft block has a very simple interpretation: it represents the lowest order interaction vertex from each operator at a given order in the derivative expansion, when all external legs are taken on-shell. Recall that in ChPT, and more generally in the NLSM, each operator at a fixed derivative order actually contains an infinite number of operators with increasing power of $1/f$. For example, the LO operator in Eq.~(\ref{eq:nlsmir}) contains all operators carrying two derivatives and powers of $1/f^2$:
\beq
{\cal L}^{(2)}= \frac{f^2}{2}{\rm tr}\left(d_\mu d^\mu\right)=  \sum_{n\ge 2} \frac{c_{2,n}}{ f^{n-2}} \partial^{2} \left[{\Pi}\right]^{n} \ ,
\label{eq:L2deriscalar}
\eeq
where  ${\Pi}= \{\pi^a\}$ and $\partial^{2} \left[{\Pi}\right]^{n}$ denotes generic contractions of $n$ $\pi^a$'s. The coefficients $c_{2,n}$ can be computed using Eq.~(\ref{eq:irdefe}), although its values are parameterization dependent.\footnote{Except for $c_{2,2}=1/2$ which is fixed by canonically normalizing the kinetic term.} There is, however, an "invariant" statement, independent of the particular parameterization chosen: the $c_{2,n}$'s must be such that the Alder's zero condition is satisfied for all S-matrix elements of ${\cal O}(p^2)$. Assuming the adjoint representation of $\SU(N)$, there is only one (flavor-ordered) soft block at ${\cal O}(p^2)$ \cite{Low:2019ynd}, up to momentum conservation,
\bea
{\cal S}^{(2)}(1,2,3,4) &=& c_0\, \frac{s_{13}}{f^2}\ ,\label{eq:o2sos}
\eea
where we have imposed the cyclic permutation invariance for the flavor-ordered partial amplitudes defined in Eq.~(\ref{eq:stfod}). ${\cal S}^{(2)}(1,2,3,4)$ is precisely the flavor-ordered 4-pt contact interaction from  Eq.~(\ref{eq:L2deriscalar}) for ChPT.\footnote{There is a second, "double-trace" soft block for Nambu-Goldstone bosons transforming under the fundamental representation of $\SO(N)$ group \cite{Low:2019ynd}.} Soft bootstrap, either through the shift symmetry in Eq.~(\ref{eq:shiftall}) or the on-shell recursion relation in Eq.~(\ref{eq:sbrr}),  uniquely determines coefficients of all higher multiplicity operators $c_{2,n}, n\ge 6$ in terms of $c_0$. Since there is only one operator at the leading order, $c_0$ can be absorbed into the normalization of $f$ and all two derivative operators are resummed into the compact form ${\rm tr}(d^\mu d_\mu)$.

Beyond LO, the effective Lagrangian has a schematic form given by
\beq
{\cal L}= \Lambda^2 f^2 \sum_{m\ge 1; n\ge 2} \frac{c_{2m,n}}{\Lambda^{2m} f^{n}} \partial^{2m} \left[{\Pi}\right]^{n} \ ,
\label{eq:Lmultiscalar}
\eeq
where the Lorentz indices on the derivatives has been suppressed and  invariance under Lorentz transformations implies an even number of derivatives.  For $(m,n)=(2,4)$, there exist four soft blocks \cite{Low:2019ynd}:
\begin{align}
& &\text{Single-trace:}\ \ {\cal S}_1^{(4)} (1,2,3,4) &=  \frac{c_1}{\Lambda^2 f^2} \  s_{13}^2 \ , \ \  & {\cal S}_2^{(4)} (1,2,3,4)&=   \frac{c_2}{\Lambda^2 f^2} \  s_{12} s_{23}\ ,\label{eq:sa4s}\\
&&\text{Double-trace:}\ \ {\cal S}_1^{(4)} (1,2|3,4) &=   \frac{d_1}{\Lambda^2 f^2}  \ s_{12}^2 \ , \ \  & {\cal S}_2^{(4)} (1,2|3,4)&=    \frac{d_1}{\Lambda^2 f^2} \ s_{13} s_{23}\ ,\label{eq:sa4d}
\end{align}
These four soft blocks can be matched to the 4-pt vertices from the four ${\cal O}(p^4)$ operators in Eq.~(\ref{eq:lecdef}). In terms of the LEC's, we have 
\bea
 c_1 =L_{4,3}+ 3 L_{4,4} ,\quad c_2 = 2(L_{4,3}- L_{4,4}),\quad d_1 &=& 2L_{4,1} + L_{4,2},\quad d_2 = 2L_{4,2}.
\eea
For each soft block, a series of operators of the form  $\partial^{4} \left[{\Pi}\right]^{n}$ are uniquely determined and resummed into the corresponding linear combination of $O_i$ in Eq.~(\ref{eq:lecdef}). The low-energy constants, or the Wilson coefficients, are  free parameters associated with each soft block during the procedure of soft bootstrap.

One important advantage of working with soft blocks, instead of operators in the Lagrangian, is they are on-shell objects free of the redundancies associated with the Lagrangian formulation. As a consequence, it is simple and straightforward to check and verify the linear dependence of two operators. For example, the equation of motion and integration-by-parts manifest themselves trivially in the on-shell approach by way of on-shell condition $p^2=0$ and momentum conservation $\sum_i p_i=0$. On the other hand, the hermiticity of an operator, which is not always transparent for higher derivative operators, can be seen through the  reflection invariance (inside each trace) of a flavor-ordered soft block, since under hermitian the $m$-trace flavor factor transforms as:
\bea
&&\left[\tr\left( T^{a_1} T^{a_2} \cdots T^{ a_{i_1}} \right)\tr\left( T^{a_{i_{1+1}}}\cdots T^{ a_{i_2}} \right)\cdots \tr\left( T^{a_{i_{m+1}}}    \cdots T^{ a_{n}} \right)\right]^\dagger \nonumber\\
&&\qquad=\tr\left(  T^{ a_{i_1}}\cdots T^{a_2} T^{a_1}  \right)\tr\left(  T^{ a_{i_2}}\cdots T^{a_{i_{1+1}}}\right)\cdots \tr\left(  T^{ a_{n}}\cdots T^{a_{i_{m+1}}} \right)\label{eq:softHermitian}
\eea
So the hermitian or anti-hermitian condition for soft blocks implies 
\bea
&& {\cal S}_n^{(m)}(1,2,\cdots, i_1|i_1+1,\cdots,i_2|\cdots|i_{m+1},\cdots,n) \nonumber \\
&&\qquad = \pm {\cal S}_n^{(m)}(i_1,\cdots, 2,1|i_2,\cdots,i_1+1|n,\cdots, i_{m+1})\ .
\eea
Furthermore, different operators related to each other under special relations such as those in Eq.~(\ref{eq:covd0}) would result in identical S-matrix elements and, therefore, give rise to identical or dependent soft blocks. These properties greatly simplify the counting of independent operators in ChPT.

Last but not least, we would like to make a comment on the basis for kinematic invariants. Since soft blocks with $2m$-derivative are homogenous polynomials of degree $m$ made out of kinematic invariants, and the number of kinematic invariants grew factorially with the number of external legs $n$, it is important to adopt an appropriate basis of independent kinematic invariant $K_{n}$.

The Mandelstam kinematic invariant basis  $K_{n}$, for $n$-external momenta $p_i, {i=1,2,3,\cdots,n}$, is defined by the symmetric invariants  $s_{ij} \equiv (p_i + p_j)^2$ and antisymmetric invariants $\epsilon_{p_1p_2\cdots p_n}\equiv\epsilon_{\mu_1\mu_2\cdots \mu_n}p_1^{\mu_1}p_2^{\mu_2}\cdots p^{\mu_n}$, where all the momenta are assumed to be incoming and $\epsilon_{\mu_1\mu_2\cdots \mu_n}$ is the antisymmetric tensor. The symmetric  invariants are even under parity transformations and antisymmetric invariants are odd under parity transformations. In the case of $n=4$ particles, the usual Mandelstam variables are defined as $s\equiv s_{12}=(p_1+p_2)^2$, $t\equiv s_{13}=(p_1+p_3)^2$, and $u\equiv s_{23}=(p_2+p_3)^2$. In this work, we are interested in computing the even-parity soft-blocks and then compare these soft-blocks with the operators that have been obtained in Refs.~\cite{Bijnens:1999sh, Bijnens:2019eze, Bijnens:2018lez}. Therefore,  we impose parity invariance which eliminates the antisymmetric invariants involving the $\epsilon$-tensor. 

Among the $s_{ij}$'s, on-shell condition and total momentum conservation together leave only $n(n-1)/2-n=n(n-3)/2$ kinematic invariants. In addition, in $d$-dimensional spacetime, there can only be at most $d$ linearly independent momenta. This implies the following $n\times n$ matrix
\begin{equation}
\begin{pmatrix} 
s_{11} & s_{12} & \dots &s_{1n} \\
\vdots & \vdots & \ddots &\vdots \\
s_{1n} & s_{2n} &\dots  & s_{nn} 
\end{pmatrix},
\label{eq:gram}
\end{equation}
has a rank of at most $d$, leading to the condition that any $(d+1)\times(d+1)$ sub-Gram matrix has zero determinant. The Gram conditions are highly nonlinear relations among $s_{ij}$.  Following Refs.~\cite{Bijnens:1999sh, Bijnens:2019eze, Bijnens:2018lez} we  will work in $d\ge n-1$ spacetime dimensions and ignore the Gram conditions. It will be interesting to return to incorporating the Gram conditions in the future. In the end, the Mandelstam basis consists of choosing any $n(n-3)/2$ invariants out of the set $\{s_{ij};i<j, i,j=1, \cdots, n\}$.

\section{Computing Soft Blocks }
\label{sec:comuting}

In ChPT, the operator at ${\cal O}\left(p^6\right)$ and ${\cal O}\left(p^8\right)$ that are hermitian, even-parity, and Lorentz invariant was constructed in  Refs.~\cite{Bijnens:1999sh, Bijnens:2019eze, Bijnens:2018lez}. In this section we will enumerate the number of soft blocks all the way up to ${\cal O}(p^{10})$ and demonstrate equivalence to the existing counting up to ${\cal O}(p^8)$. In the following subsections, we compute the $4$-pt, $6$-pt, $8$-pt, and $10$-pt soft-blocks up to order ${\cal O}(p^{10})$ in the momentum space.

\begin{table}[t!]
\begin{tabular}{|c|c|c|c|}
\hline
\begin{tabular}[c]{@{}c@{}} Order \\ ${\cal O}\left(p^{2m}\right)$\end{tabular}             & \begin{tabular}[c]{@{}c@{}} External Legs \\ $n$-pt \end{tabular} & \begin{tabular}[c]{@{}l@{}}$N$ before Adler's zero\end{tabular} & \begin{tabular}[c]{@{}l@{}} $N$ after Adler's zero\end{tabular} \\ \hline
${\cal O}\left(p^4\right)$                  &   $4$-pt            &  $3$          & $3$  \\ \hline
\multirow{2}{*}{${\cal O}\left(p^6\right)$} &      $4$-pt               &   $4$ & $4$ \\ \cline{2-4} 
                  &   $6$-pt                   &  $165$ & $15$                                                                \\ \hline
\multirow{3}{*}{${\cal O}\left(p^8\right)$} &  $4$-pt                   &  $5$  &$5$                                                                            \\ \cline{2-4} 
                  &  $6$-pt                   & $495$   &   $150$   
                  \\ \cline{2-4} 
                  &  $8$-pt                  &  $8855$    & $105$                                                                            \\ \hline
\multirow{4}{*}{${\cal O}\left(p^{10}\right)$} &    $4$-pt   &  $6$ &$6$                                                                           \\ \cline{2-4} 
                  &  $6$-pt                & $1287$     & $621$                                                                            \\ \cline{2-4} 
                  &$8$-pt               &  $42504$  & $3360$                                                                      \\ \cline{2-4} 
                  & $10$-pt         & $575757$      & $945$                                                                           \\ \hline
\end{tabular}
\caption{The number of monomials $N$ before and after imposing the Adler's zero condition up to ${\cal O}\left(p^{10}\right)$.
}
\label{tab:Adler}
\end{table}

For $n$-pt soft blocks, we choose the set of independent kinematic invariants in the Mandelstam basis to be 
\begin{equation}
\{s_{12},s_{13}, \cdots, s_{1,n-1}, s_{23},s_{24}, \cdots, s_{2,n-1}, \cdots, s_{n-3,n-2}, s_{n-3,n-1} \}.
\label{eq:kinematic_basis}
\end{equation}
In order to construct the $n$-pt soft blocks at ${\cal O}\left(p^{2m}\right)$ in momenta, 
we start with the most general homogeneous polynomial of degree $m$ in the kinematic invariants in $K_n$,
\begin{equation}
p[s_{ij}]=c_1 s_{12}^m + c_1 s_{12}^{m-1} s_{13}+\cdots+c_N s_{n-3,n-1}^m,
\label{eq:poly}
\end{equation}
where the number of coefficients (number of monomials) in the polynomial $p[s_{ij}]$ is
\begin{equation}
N= \binom{K_n+m-1}{m}\ .
\label{eq:polynum}
\end{equation}
The coefficients $\{c_1, c_2, \cdots, c_N\}$ of the polynomial $p[s_{ij}]$ in Eq.~\eqref{eq:poly} must satisfy constraints from the Adler's zero condition, invariance under cyclic permutations and hermitian conjugation. To give an idea of the vast number of monomials we are dealing with, we show in Table~\ref{tab:Adler} the number of monomials $N$  before and after imposing the Adler zero condition on every external leg. 

After the Adler's zero condition, we still need to impose the invariance under cyclic permutations and hermitian conjugation, only  after then do we arrive at the number of independent $n$-pt soft blocks at ${\cal O}(p^{2m})$. In the following subsections, we explicitly construct the $n$-pt soft-blocks up to ${\cal O}(p^{10})$ in a general spacetime dimension $d$. The $4$-pt soft-blocks up to  ${\cal O}(p^{4})$ have already been studied in Ref.~\cite{Low:2019ynd}. 

\subsection{$4$-pt Soft Blocks}
\label{subsec:4pt} 
In the subsection, we compute the 4-pt soft-blocks up to  ${\cal O}(p^{10})$. As an illustration, we explain in detail how to compute the $4$-pt soft-block at  ${\cal O}\left(p^6\right)$.

At $4$-pt, there are two independent kinematic invariants, $\left\{s_{12},s_{13}\right\}$.  
 In order to compute the $4$-pt soft blocks,  at  ${\cal O}\left(p^6\right)$, we start with a general homogeneous polynomial of degree-3 in $K_4=\left\{s_{12},s_{13}\right\}$:
\begin{equation}
p[s_{ij}]=c_0 s_{12}^3+c_1 s_{12}^2 s_{13}+c_2 s_{12} s_{13}^2+c_3 s_{13}^3.
\label{eq:poly4pt}
\end{equation}
It turned out that, because of total momentum conservation and on-shell condition, the above equation automatically satisfy the Adler's zero condition for all external momentum. Consequently, we only need to consider the invariance of Eq.~\eqref{eq:poly4pt} under cyclic symmetry and hermitian conjugation. 

It is easy to see that at $4$-pt there are only two  possible trace structures, which already appear at ${\cal O}(p^4)$ \cite{Low:2019ynd}: i) the single trace soft block ${\cal S}^{(6)}(1,2,3,4)$, where $(1,2,3,4)$ indicates invariance under the cyclic transformation  in $(1234)$, and ii) the double-trace soft block ${\cal S}^{(6)}(1,2|3,4)$, where $(1,2|3,4)$ indicates the separate invariance under cyclic transformations $(12)$ and $(34)$.  For the double-trace soft block there is an additional symmetry under exchange of the two traces, $ (12)\leftrightarrow(34)$.

Let's consider the single trace soft-block  ${\cal S}^{(6)}(1,2,3,4)$ first. The polynomial $p[s_{ij}]$ in Eq.~\eqref{eq:poly4pt} under the cyclic transformation in $(1234)$ becomes
\begin{eqnarray}
p[s_{ij}] \to p'[s_{ij}] &=&c_0 s_{23}^3+c_1 s_{23}^2 s_{24}+c_2 s_{23} s_{24}^2+c_3 s_{24}^3\nonumber\\
&=&c_0 (-s_{12}-s_{13})^3+c_1 ({-s_{12}-s_{13}})^2 s_{13}+c_2 ({-s_{12}-s_{13}}) s_{13}^2+c_3 s_{13}^3.
\end{eqnarray}
Note that the cyclic transformation takes us outside of $K_4=\{s_{12}, s_{13}\}$ and we have used momentum conservation to re-express $p'[s_{ij}]$ in $K_4$. Setting $p[s_{ij}] = p'[s_{ij}]$, as required by the cyclic invariance, we have
\begin{equation}
p[s_{ij}]=c_0\, s_{13}^3+ c_1\, s_{12} s_{13} (s_{12} + s_{13}).
\label{eq:polyst}
\end{equation}
So only two independent coefficients remain, giving rise to two single trace soft blocks:
\begin{align}
& & {\cal S}_1^{(6)} (1,2,3,4) &=  \frac{c_0}{\Lambda^4 f^2} \  s_{13}^3 \ , \ \  & {\cal S}_2^{(6)} (1,2,3,4)&=   \frac{c_1}{\Lambda^4 f^2} \  s_{12} s_{13} (s_{12} + s_{13})\ ,\label{eq:4st6}
\end{align}
where $c_0$ and $c_1$ are related to the ${\cal O}(p^6)$ LEC's, with the precise relations depending on the particular nonlinear parameterization chosen.
It is easy to check that both soft blocks  in Eq.\eqref{eq:4st6} are invariant under the hermitian conjugation transformation: $(1234) \to (4321)$, although we will see later that this is not always automatic. 

Similarly, the two independent double-trace soft blocks ${\cal S}^{(6)}(12|34)$, which are invariant under the individual cyclic permutations  $(12)$ and $(34)$, exchange symmetry of the trace  $(12)\leftrightarrow(34)$, and hermitian conjugation $(12)\to(21)\ ,(34)\to(43)$ are
\begin{align}
& & {\cal S}_1^{(6)} (1,2|3,4) &=  \frac{d_0}{\Lambda^4 f^2} \  s_{12}^3 \ , \ \  & {\cal S}_2^{(6)} (1,2|3,4)&=   \frac{d_1}{\Lambda^4 f^2} \  s_{12} s_{13} (s_{12} + s_{13})\ ,\label{eq:4dt6}
\end{align}
where $d_0$ and $d_1$ are another two LEC's.

Following the above procedures, we compute the $4$-pt soft-blocks up to ${\cal O}(p^{10})$. Single trace soft blocks for 4-pt at ${\cal O}(p^{8})$ are
\begin{align}
	& & {\cal S}_1^{(8)} (1,2,3,4) &=  \frac{c_0'}{\Lambda^6 f^2} \,  s_{13}^4\, ,  \nonumber \\
	&&  {\cal S}_2^{(8)} (1,2,3,4) & =\frac{c_1'}{\Lambda^6 f^2} \, s_{12}^2 \left(s_{12}+s_{13}\right)^2\, , \nonumber \\
	&& {\cal S}_3^{(8)} (1,2,3,4)&=   \frac{c_2'}{\Lambda^6 f^2}\, s_{12} \left(s_{13}^3-2s_{13}s_{12}^2-s_{12}^3\right)\, , \label{eq:4ptp8single}
\end{align}
where $c_0'$, $c_1'$, and $c_2'$ are three LEC's. The 4-pt double-trace soft blocks at ${\cal O}(p^{8})$ are
\begin{align}
	& & {\cal S}_1^{(8)} (1,2|3,4) &=  \frac{d_0'}{\Lambda^6 f^2}\,  s_{12}^4 \ ,  \nonumber\\
	&&  {\cal S}_2^{(8)} (1,2|3,4) &=   \frac{d_1'}{\Lambda^6 f^2} \, s_{12}^2 s_{13} (s_{12} + s_{13})\, , \nonumber\\
	&& {\cal S}_3^{(8)} (1,2|3,4)&=   \frac{d_2'}{\Lambda^6 f^2} \, s_{13}\left(s_{13}^3+2 s_{12} s_{13}^2-s_{12}^3 \right)\, ,\label{eq:4ptp8double} 
\end{align}
where $d_0'$, $d_1'$, and $d_2'$ are three LEC's. The 4-pt single trace soft blocks for at ${\cal O}(p^{10})$ are
\begin{align}
	& & {\cal S}_1^{(10)} (1,2,3,4) &=  \frac{c_0''}{\Lambda^8 f^2} \,  s_{13}^5\, , \nonumber\\
	&& {\cal S}_2^{(10)} (1,2,3,4)&=   \frac{c_1''}{\Lambda^8 f^2} \, s_{12}^2 s_{13} \left(s_{12}+s_{13}\right)^2 \, , \nonumber \\
	&& {\cal S}_3^{(10)} (1,2,3,4)&=\frac{c_2''}{\Lambda^8 f^2} \, s_{12} s_{13} \left(s_{13}^3-2 s_{13} s_{12}^2  -s_{12}^3\right)\, , \label{eq:4ptp8single}
\end{align}
where $c_0''$, $c_1''$, and $c_2''$ are two LEC's, Tge 4-pt double-trace soft blocks at ${\cal O}(p^{10})$ are
\begin{align}
	& & {\cal S}_1^{(10)} (1,2|3,4) &=  \frac{d_0''}{\Lambda^8 f^2} \, s_{12}^5 \ ,  \nonumber\\
	&&  {\cal S}_2^{(10)} (1,2|3,4)&=   \frac{d_1''}{\Lambda^8 f^2} \, s_{12}^3 s_{13} (s_{12} + s_{13})\, , \nonumber\\
	&& {\cal S}_3^{(10)} (1,2|3,4)&=   \frac{d_2''}{\Lambda^8 f^2} \, s_{12} s_{13} \left(s_{13}^3+2 s_{12} s_{13}^2 -s_{12}^3\right)\, .\label{eq:4ptp8double} 
\end{align}
where $d_0''$, $d_1''$, and $d_2''$ are three LEC's. Note that, for 4-pt, since $s_{13}$ is cyclic symmetric for the single trace and $s_{12}$ is cyclic symmetric for the double trace, ${\cal O}(p^{10})$ soft blocks for single and double trace can be obtained from soft blocks at ${\cal O}(p^{8})$ by multiplying $s_{13}$ and $s_{12}$ respectively. This feature is also true going from ${\cal O}(p^4)$ to ${\cal O}(p^6)$ for 4-pt; but for higher point soft blocks, these features disappear. In Table \ref{tab:4pt} we list the number of LEC;s for 4-pt soft blocks up to ${\cal O}(p^{10})$. 

\begin{table}[t!]
	\begin{center}
		\begin{tabular}{|c|c|c|}
			\hline
			& $(1234)$ & $(12|34)$\\ 
			\hline
			${\cal O}(p^4)$  & $2$ & $2$\\
			\hline
			${\cal O}(p^6)$  & $2$ & $2$\\
			\hline
			${\cal O}(p^8)$  & $3$ & $3$\\
			\hline
			${\cal O}(p^{10})$  & $3$ & $3$\\
			\hline	
		\end{tabular}
	\end{center}
	\caption{The number of independent $4$-pt soft blocks up to ${\cal O}(p^{10})$.}
	\label{tab:4pt}
\end{table}%

\subsection{$6$-pt Soft Blocks}
\label{subsec:6pt}
At $6$-pt or higher, the number of monomials in Eq.~\ref{eq:poly} grows factorially, as  shown in Table~\ref{tab:Adler}. 
Moreover, the Adler's zero condition is not automatic anymore and now gives non-trivial constraints on the coefficients $c_i$ in $p[s_{ij}]$, which significantly reduces the number of monomials.

The kinematic invariants we adopt at 6-pt are:
\begin{equation}
K_6=\{s_{12},s_{13},s_{14},s_{15},s_{23},s_{24},s_{25},s_{34},s_{35}\}\ .
\label{eq:6pt_kinematic}
\end{equation}
The Adler's zero condition corresponds to taking the soft limit for every external momentum. For example, let's take $p_1\to 0$. This imposes constraints on four of the kinematic invariants in $K_6$, thereby reducing the number of independent kinematic invariants down to five. It is obvious that the same is true for taking any other $p_i \to 0$. More explicitly,
\begin{eqnarray}
\label{eq:6ptAdler}
p_1 \to 0\ \ &\Longrightarrow&\ \ \{s_{12}\to 0,s_{13} \to 0, s_{14} \to 0,s_{15}\to 0\}, \nonumber\\
p_2 \to 0\ \ &\Longrightarrow&\ \ \left\{s_{12}\to 0,s_{23}\to 0,s_{24}\to 0,s_{25}\to 0\right\}, \nonumber\\
p_3 \to 0\ \ &\Longrightarrow&\ \ \left\{s_{13}\to 0,s_{23}\to 0,s_{34}\to 0,s_{35}\to 0\right\}, \nonumber\\
p_4 \to 0\ \ &\Longrightarrow&\ \ \left\{s_{14}\to 0,s_{24}\to 0,s_{34}\to 0,s_{35}\to -s_{12}-s_{13}-s_{15}-s_{23}-s_{25}\right\}, \nonumber\\
p_5 \to 0\ \ &\Longrightarrow&\ \ \left\{s_{15}\to 0,s_{25}\to 0,s_{35}\to 0,s_{34}\to -s_{12}-s_{13}-s_{14}-s_{23}-s_{24}\right\}, \nonumber\\
p_6 \to 0\ \ &\Longrightarrow&\ \ \left\{s_{15}\to -s_{12}-s_{13}-s_{14},s_{25}\to -s_{12}-s_{23}-s_{24},\right. \nonumber\\
&&\ \  \ \ \left. s_{34}\to -s_{12}-s_{13}-s_{14}-s_{23}-s_{24},s_{35}\to s_{12}+s_{14}+s_{24}\right\}.
\end{eqnarray}
These are the substitutions one performs on $p[s_{ij}]$ in Eq.~(\ref{eq:poly}) when taking the soft limits. Requiring $p[s_{ij}]=0$ under the soft limits greatly reduces the number of monomials.
For instance, in the case of 6-pt at ${\cal O}(p^{6})$, the Adler's zero condition imposes 150  constraints on $c_i$'s, and thus reduces the number of monomials from 165  to 15 as shown in Table~\ref{tab:Adler}. For completeness we provide the explicit expressions of  $p[s_{ij}]$ in this case, after imposing the Adler's zero condition, in Eq.~\eqref{eqapp:6ptAdler}.

\begin{table}[htp]
	\begin{center}
		\begin{tabular}{|c|c|c|c|c|}
			\hline
			Trace structures & $(123456)$ & $(12|3456)$ & $(123|456)$ & $(12|34|56)$ \\ 
			\hline
			${\cal O}(p^6)$ & $5\,(0)$ & $4\,(0)$ & $3\,(0)$ & $3\,(0)$\\
			\hline
			${\cal O}(p^8)$ & $22\,(7)$ & $19\,(6)$ & $10\,(3)$ & $9\,(0)$\\
			\hline
			${\cal O}(p^{10})$ & $71\,(41)$ & $59\,(32)$ & $28\,(15)$ & $25\,(0)$\\
			\hline
		\end{tabular}
	\end{center}
	\caption{The number of $6$-pt soft blocks up to ${\cal O}(p^{10})$. The numbers in the parentheses are the numbers of non-Hermitian soft blocks.}
	\label{tab:6pt}
\end{table}%

After Adler's zero condition we need to consider the possible trace structures at 6-pt. There are four possibilities listed in Table~\ref{tab:6pt} and we need to impose the corresponding cyclic invariance, exchange symmetry of the trace, as well as hermitian conjugation. In the end we found all 6-pt soft blocks up to ${\cal O}(p^{10})$, whose numbers are summarized in  Table~\ref{tab:6pt}. Explicit expressions for the ${\cal O}\left(p^6\right)$ $6$-pt soft blocks are given in Eqs.~\eqref{eqapp:6ptst1}--\eqref{eqapp:6pttt2} in appendix~\ref{app:6pt}. At ${\cal O}\left(p^8\right)$ and ${\cal O}\left(p^{10}\right)$ the expressions are quite complicated already for 6-pt soft blocks. The results are given instead in the accompanying {\tt Mathematica} notebook, which contain the complete soft blocks we computed up to ${\cal O}(p^{10})$.

\subsection{$8$-pt and $10$-pt Soft Blocks}
\label{subsec:8-10pt}
We follow the procedures mentioned in sections~\ref{subsec:4pt} and \ref{subsec:6pt} to compute the $8$-pt and $10$-pt soft blocks up to ${\cal O}\left(p^{10}\right)$. For $8$-pt, there are seven possibilities for the different trace structures as listed in Table~\ref{tab:8pt}, while for $10$-pt there are twelve possibilities  listed in Table~\ref{tab:10pt}. We impose the cyclic invariance, trace exchange symmetry of the trace and the hermitian conjugation corresponding to the trace structure. The computation of the soft blocks is done in the accompanying {\tt Mathematica} notebook. Here we list the number of soft blocks for $8$-pt and $10$-pt in Table
\ref{tab:8pt} and Table \ref{tab:10pt}, respectively. The number of of soft blocks up to ${\cal O}\left(p^8\right)$ that satisfy the Hermitian (CP even) condition matches exactly the number of independent operators for $\SU(N)$ ChPT obtained in Refs.~\cite{Bijnens:1999sh,  Bijnens:2018lez,Bijnens:2019eze}.
\begin{table}[t!]
	\begin{center}
		\begin{tabular}{|c|c|c|c|c|c|c|c|}
			\hline
			& $(12345678)$ & $(12|345678)$ & $(123|45678)$ & $(1234|5678)$ & $(12|34|5678)$ & $(12|345|678)$ & $(12|34|56|78)$\\ 
			\hline
			${\cal O}(p^{8})$ & $17\,(1)$ & $13\,(1)$ & $7\,(0)$ & $10\,(0)$ & $10\,(0)$ & $7\,(0)$ & $5\,(0)$\\
			\hline
			${\cal O}(p^{10})$ & $255\,(180)$ & $190\,(130)$ & $128\,(98)$ & $87\,(42)$ & $107\,(42)$ & $76\,(41)$ & $26\,(0)$\\
			\hline
		\end{tabular}
	\end{center}
	\caption{The number of $8$-pt soft blocks up to ${\cal O}(p^{10})$. The numbers in the parentheses are the numbers of non-Hermitian soft blocks.}
	\label{tab:8pt}
\end{table}
\begin{table}[t!]
	\begin{center}
		\begin{tabular}{|c|c|c|c|c|c|c|}
			\hline
			& $(0123456789)$ & $(01|23456789)$ & $(012|3456789)$ & $(0123|456789)$ & $(01234|56789)$ & $(01|23|456789)$ \\ 
			\hline
			${\cal O}(p^{10})$ & $79\,(26)$ & $56\,(18)$ & $35\,(10)$ & 44\,(6)  & 25\,(4)  & 34\,(3) \\
			\hline
			\hline
			& $(01|234|56789)$ & $(01|2345|6789)$ & $(012|345|6789)$ & $(01|23|45|6789)$ & $(01|23|456|789)$ & $(01|23|45|67|89)$ \\ 
			\hline
			${\cal O}(p^{10})$ & 29\,(6) & 27\,(3) & 21\,(0) & 18\,(0) & 18\,(0) & 7\,(0)\\
			\hline 
		\end{tabular}
	\end{center}
	\caption{The number of $10$-pt soft blocks at ${\cal O}(p^{10})$. The numbers in the parentheses are the numbers of non-Hermitian soft blocks.}
	\label{tab:10pt}
\end{table}

\section{Summary}
\label{sect:summary}

In this work we computed the number of independent soft blocks in ChPT up to N$^4$LO at ${\cal O}(p^{10})$, in the massless quark limit and with all external sources turned off. In particular, to facilitate comparison with existing literature, we assumed a general spacetime dimension $d$ and general flavor $N_f$. The soft blocks are in one-to-one correspondence with the independent operators, and therefore the LEC's, in ChPT. In Table \ref{tab:summary} we present a summary of the total number of soft blocks up to ${\cal O}(p^{10})$. The complete effective Lagrangian of ChPT at ${\cal O}(p^6)$ and ${\cal O}(p^8)$ are known in Refs.~\cite{Gasser:1983yg,Fearing:1994ga,Bijnens:1999sh,Bijnens:2018lez}. Our "on-shell" approach is independent from the traditional Lagrangian formulation utilizing the CCWZ method, and comparison of our results with the literature finds agreements after taking the appropriate limit. In particular, we make a prediction at ${\cal O}(p^{10})$.

In addition to counting the soft blocks, we also presented the expressions for the soft block themselves. These are important ingredients for pursuing the soft bootstrap program, which aims to recursively construct on-shell tree amplitudes of ChPT. The soft blocks are connected to the Lagrangian approach  as the lowest pt contact interaction in each of the invariant operators, when all external legs are taken on-shell. As on-shell objects, they are  free of ambiguities arising from operator relations typically present in the Lagrangian formulation.

One limitation of the current study is we are only considering ChPT in the massless quark limit and with external sources turned off, which is due to the fact that the soft bootstrap approach has not been extended to these cases. It would certainly be very interesting to study how this could be done in the future.

Although it is unlikely that results presented here could be of immediate phenomenological interest, it is perhaps worth recalling that ChPT has been playing a prominent role in our understanding of low-energy QCD in the past half-a-century, and also is the prototypical example of effective field theories. Thus it is rather remarkable that new underlying structures are still being uncovered for such an important and well-studied subject. In this context it will be useful to compare and understand different recent approaches in enumerating operators in EFT's.

\begin{table}[t!]
	\begin{center}
		\begin{tabular}{|c|c|c|c|c|}
			\hline
			 & ${\cal O}(p^4)$ & ${\cal O}(p^6)$ & ${\cal O}(p^8)$ & ${\cal O}(p^{10})$ \\ 
			\hline
			No. of Soft Blocks& 4 &19 & 135 & 1451\\
			\hline
		\end{tabular}
	\end{center}
	\caption{A summary table for the number of soft blocks, including all trace structures.}
	\label{tab:summary}
\end{table}%

\acknowledgements
IL is supported in part by the U.S. Department of Energy under contracts No. DE- AC02-06CH11357 at Argonne and No. DE-SC0010143 at Northwestern.
LD, TM and AM are supported by in part by Director, Office of Science, Office of Nuclear Physics, of the U.S. Department of Energy under grant number DE-FG02-05ER41368. LD and TM are also supported in part by the U.S. Department of Energy, Office of Science, Office of Nuclear Physics, within the framework of the TMD Topical Collaboration.

\appendix
\section{$6$-pt polynomial satisfying Adler's zero condition}
\label{app:6ptAdler}
The general polynomial $p[s_{ij}]$ in Eq.~\eqref{eq:poly} for $6$-pt at ${\cal O}\left(p^6\right)$ has $165$ coefficients $c_i$ to begin with. The implementation of the Adler's zero condition in Eq.~\eqref{eq:6ptAdler} leads to $150$ constraints. Thus, the polynomial $p[s_{ij}]$ in Eq.~\eqref{eq:poly} reduces to
\begin{align}
 p[s_{ij}]&= c_0 \Big[s_{14} s_{23} \left(s_{12}+s_{13}+s_{14}+s_{23}+s_{24}+s_{34}\right)\Big]\nonumber\\
 &+c_1 \Big[s_{13} s_{24} \left(s_{12}+s_{13}+s_{14}+s_{23}+s_{24}+s_{34}\right)\Big]\nonumber\\
 &+c_2 \Big[s_{12} s_{34} \left(s_{12}+s_{13}+s_{14}+s_{23}+s_{24}+s_{34}\right)\Big]\nonumber\\
 &+c_3\Big[\left(s_{12}+s_{13}+s_{14}+s_{15}\right) s_{24} s_{35}\Big]\nonumber\\
 &+c_4\Big[s_{14} \left(s_{12}+s_{23}+s_{24}+s_{25}\right) s_{35}\Big]\nonumber\\
 &+c_5\Big[s_{15} s_{23} \left(s_{12}+s_{13}+s_{15}+s_{23}+s_{25}+s_{35}\right)\Big]\nonumber\\&+c_6\Big[s_{13} s_{25} \left(s_{12}+s_{13}+s_{15}+s_{23}+s_{25}+s_{35}\right)\Big]\nonumber\\
 &+c_7\Big[s_{12} s_{35} \left(s_{12}+s_{13}+s_{15}+s_{23}+s_{25}+s_{35}\right)\Big]\nonumber\\
 &+c_8\Big[s_{14} s_{25} \left(s_{13}+s_{23}+s_{34}\right)-s_{14} \left(s_{12}+s_{23}+s_{24}\right) s_{35}\Big]\nonumber\\
 &+\frac{c_9}{2}\Bigg[ s_{12} \bigg(s_{13}^2+\left(s_{14}+s_{15}+2 s_{23}+s_{24}+s_{25}+s_{34}+s_{35}\right) s_{13}+s_{23}^2\nonumber\\&+s_{14} s_{23}+s_{15} s_{23}+s_{12} \left(s_{13}+s_{23}\right)+s_{23} s_{24}+s_{23} s_{25}+s_{15} s_{34}+s_{23} s_{34}\nonumber\\&+s_{25} s_{34}+\left(s_{14}+s_{23}+s_{24}+2 s_{34}\right) s_{35}\bigg)\Bigg]\nonumber\\
 &+c_{10}\Bigg[s_{13} \bigg(s_{12}^2+\left(s_{13}+s_{14}+s_{15}+2 s_{23}+s_{24}+s_{25}+s_{34}+s_{35}\right) s_{12}+s_{23}^2\nonumber\\&+s_{13} s_{23}+s_{14} s_{23}+s_{15} s_{23}+s_{15} s_{24}+s_{23} s_{24}+s_{14} s_{25}+s_{23} s_{25}+2 s_{24} s_{25}\nonumber\\&+s_{23} s_{34}+s_{25} s_{34}+\left(s_{23}+s_{24}\right) s_{35}\bigg)\Bigg]\nonumber\\
 &+c_{11}\Bigg[\bigg(\frac{1}{2} \left(s_{23}-s_{13}\right) s_{12}^2+\frac{1}{2} \big(-s_{13}^2-\left(s_{14}+s_{15}+s_{24}+s_{25}+s_{34}+s_{35}\right) s_{13}+s_{23}^2\nonumber\\&+s_{14} s_{23}+s_{15} s_{23}+s_{23} s_{24}+s_{23} s_{25}-s_{15} s_{34}+s_{23} s_{34}+s_{25} s_{34}+\left(s_{14}+s_{23}-s_{24}\right) s_{35}\big) s_{12}\nonumber\\&+\left(s_{14}+s_{24}\right) \left(s_{15} s_{23}-s_{13} s_{25}\right)+\left(s_{14} s_{23}-s_{13} s_{24}\right) s_{35}\bigg)\Bigg]\nonumber\\ &+c_{12}\Bigg[\bigg(\frac{1}{2} \left(s_{13}-s_{23}\right) s_{12}^2+\frac{1}{2} \big(s_{13}^2+\left(s_{14}+s_{15}+s_{24}+s_{25}+s_{34}+s_{35}\right) s_{13}-s_{23}^2\nonumber\\&-s_{14} s_{23}-s_{15} s_{23}-s_{23} s_{24}-s_{23} s_{25}+s_{15} s_{34}-s_{23} s_{34}-s_{25} s_{34}-\left(s_{14}+s_{23}+s_{24}\right) s_{35}\big) s_{12}\nonumber\\&+s_{13} s_{24} \left(s_{15}+s_{25}\right)+s_{15} s_{24} s_{34}-s_{14} \left(s_{15} s_{23}-s_{13} s_{25}+\left(s_{23}+s_{24}\right) s_{35}\right)\bigg)\Bigg]\nonumber\\
 &+c_{13}\Bigg[\bigg(-\left(s_{13}^2+\left(s_{14}+s_{15}+2 s_{23}+s_{24}+s_{25}+s_{34}+s_{35}\right) s_{13}-s_{25} s_{34}-s_{14} s_{35}\right) s_{12}\nonumber\\&-s_{12}^2 s_{13}-s_{13}^2 s_{23}-s_{14} s_{23} s_{25}+s_{15} s_{25} s_{34}+s_{14} s_{23} s_{35}+s_{14} s_{24} s_{35}-s_{13} \big(s_{23}^2+s_{24} s_{23}\nonumber\\&+s_{25} s_{23}+s_{34} s_{23}+s_{15} \left(s_{23}+s_{24}\right)+2 s_{24} s_{25}+s_{14} \left(s_{23}+2 s_{25}\right)+\left(s_{23}+s_{24}\right) s_{35}\big)\bigg)\Bigg]\nonumber\\
 &+c_{14}\Bigg[\bigg(\frac{1}{2} \left(s_{13}+s_{23}\right) s_{12}^2+\frac{1}{2} \big(s_{13}^2+\left(s_{14}+s_{15}+4 s_{23}+s_{24}+s_{25}+s_{34}+s_{35}\right) s_{13}+s_{23}^2\nonumber\\&+s_{14} s_{23}+s_{15} s_{23}+s_{23} s_{24}+s_{23} s_{25}+s_{15} s_{34}+s_{23} s_{34}-s_{25} s_{34}+\left(-s_{14}+s_{23}+s_{24}\right) s_{35}\big) s_{12}\nonumber\\&+s_{13}^2 s_{23}+s_{23} \left(s_{14} \left(s_{15}+s_{25}\right)+s_{15} s_{34}\right)+s_{13} \big(s_{23}^2+s_{15} s_{23}+s_{24} s_{23}+s_{25} s_{23}+s_{34} s_{23}\nonumber\\&+s_{24} s_{25}+s_{14} \left(s_{23}+s_{25}\right)+\left(s_{23}+s_{24}\right) s_{35}\big)\bigg)\Bigg],
 \label{eqapp:6ptAdler}
 \end{align}
 where $\{c_0,c_1,\cdots,c_{14}\}$ are the $15$ coefficients remaining. This is the starting point of imposing cyclic invariance for various trace structures to compute the soft blocks.

\section{$6$-pt Soft Blocks at ${\cal O}\left(p^6\right)$}
\label{app:6pt}

Here we explicitly list the $6$-pt soft blocks at ${\cal O}\left(p^6\right)$ for the single trace, double-trace, and triple-trace structures shown in Table~\ref{tab:6pt}, after imposing Adler's zero condition, trace exchange symmetry, and the hermitian conjugation. The explicit expressions for the $6$-pt soft blocks at ${\cal O}\left(p^8\right)$ and ${\cal O}\left(p^{10}\right)$ are too long to be presented here and can found in the accompanying {\tt Mathematica} notebook.

\subsection{Single Trace}
\label{app:6ptst}
There are five single trace $6$-pt soft blocks at ${\cal O}\left(p^6\right)$ that are invariant under the cyclic transformation $\left(123456\right)$ and hermitian conjugation $(123456) \leftrightarrow (654321)$:
\begin{align}
 {\cal S}_1^{(6)} (1,2,3,4,5,6)&= \frac{a_0}{\Lambda^4 f^4}\, s_{14} s_{25} \left(s_{13}+s_{23}+s_{34}+s_{35}\right)\label{eqapp:6ptst1},\\
 {\cal S}_2^{(6)} (1,2,3,4,5,6)&= \frac{a_1}{\Lambda^4 f^4}\Big[ s_{13}s_{15}\left(s_{24}-s_{25}\right)  -s_{13}^2s_{25} -s_{13}s_{25} \left(s_{12}+s_{14}+s_{23}+s_{25}+s_{35}\right)\nonumber\\
 &-s_{14} s_{25} \left(s_{23}+s_{34}\right)+s_{14} s_{35} \left(s_{12}+s_{23}+s_{24}\right)+s_{15} s_{24} \left(s_{23}+s_{34}+s_{35}\right)\Big],\\
 {\cal S}_3^{(6)} (1,2,3,4,5,6)&= \frac{a_2}{\Lambda^4 f^4}\Bigg[s_{13}^2\left(s_{23}+s_{24}\right) +s_{13}\bigg(s_{23} s_{35}+s_{14} \left(s_{23}+s_{24}\right)+s_{15} \left(2 s_{23}+s_{24}\right)\nonumber\\
 &+\left(s_{23}+s_{24}+s_{25}\right) \left(s_{23}+s_{24}+s_{34}\right)\bigg) +s_{12} \bigg(s_{13} \left(s_{23}+s_{24}\right)+s_{15} \left(s_{23}-s_{34}\right)\nonumber\\
 &-\left(s_{24}+s_{34}\right) s_{35}\bigg)+s_{23} \bigg(s_{14} \left(s_{15}+s_{35}\right)+s_{15} \left(s_{15}+s_{23}+s_{24}+s_{25}+s_{35}\right)\bigg)\Bigg],\\
 {\cal S}_4^{(6)} (1,2,3,4,5,6)&= \frac{a_3}{\Lambda^4 f^4}\Bigg[ s_{12}^2\left(s_{23}+s_{34}\right)+s_{12}\bigg(s_{23}^2+2 s_{14} s_{23}+s_{15} s_{23}+s_{24} s_{23}+s_{25} s_{23}\nonumber\\
 &+2 s_{34} s_{23}+s_{34}^2+s_{14} s_{34}+s_{15} s_{34}+s_{24} s_{34}+s_{13} \left(s_{23}-s_{24}+s_{34}\right)\nonumber\\
 &+s_{35}\left(s_{23}+s_{24}+s_{34}\right) \bigg)+\left(s_{14} s_{23}-s_{13} s_{24}\right) \left(s_{13}+s_{14}+s_{15}+s_{23}+s_{24}+s_{25}\right)\nonumber\\
 &+ s_{23} s_{34}\left(s_{14}+s_{15}\right)-s_{13}s_{34} \left(s_{24}+s_{25}\right) \Bigg],
 \label{eqapp:6ptst61} 
 \end{align}
 \begin{align}
{\cal S}_5^{(6)} (1,2,3,4,5,6)&= \frac{a_4}{\Lambda^4 f^4}\Bigg[-s_{14} s_{23} s_{34}-s_{14} s_{23} \left(s_{13}+s_{14}+s_{23}+s_{24}\right)\nonumber\\
&+ s_{25} s_{34}\left(s_{13}+s_{14}+s_{15}\right)
- s_{12}^2\left(s_{13}+s_{23}+s_{34}+s_{35}\right)\nonumber\\
&-s_{12}\bigg(s_{13}^2+s_{23}^2+s_{34}^2+s_{35}^2+s_{15} s_{23}+s_{23} s_{24}+s_{23} s_{25}+s_{24} s_{34}+s_{14} s_{34}\nonumber\\
&+s_{15} s_{34}+2 s_{14} s_{23}+2 s_{23} s_{34}+ s_{13}\big(s_{14}+s_{15}+2 s_{23}+s_{24}
+s_{25}+2 \left(s_{34}+s_{35}\right)\big)\nonumber\\
&+\left(s_{14}+s_{15}+2 s_{23}+s_{24}+s_{25}+2 s_{34}\right)\bigg)\Bigg],
  \label{eqapp:6ptst62}
\end{align}
where $a_0$, $a_1$, $a_2$, $a_3$, and $a_4$ are arbitrary constants.
\subsection{Double-Trace}
\label{app:6ptdt}
There are two trace structures for the double-trace soft blocks: ${\cal S}^{(6)} (1,2\,|\,3,4,5,6)$ and ${\cal S}^{(6)} (1,2,3\,|\,4,5,6)$. We list the expressions of the two trace structures separately. 
\subsubsection{$\bf{\left(1,2\,|\,3,4,5,6\right)}$}
\label{app:6ptdt1}
The double-trace soft blocks ${\cal S}^{(6)} (1,2\,|\,3,4,5,6)$ are invariant under the separate cyclic transformations, $\left(12\right)$ and $\left(3456\right)$,  and hermitian conjugation: $(12)\leftrightarrow(21)$ and $(3456)\leftrightarrow(6543)$. There are four double-trace $6$-pt soft blocks  ${\cal S}^{(6)} (1,2\,|\,3,4,5,6)$ at ${\cal O}\left(p^6\right)$ that satisfy the above symmetries:
\begin{align}
 {\cal S}_1^{(6)} (1,2\,|\,3,4,5,6)&= \frac{b_0}{\Lambda^4 f^4}\, s_{12} s_{35} \left(s_{12}+s_{13}+s_{15}+s_{23}+s_{25}+s_{35}\right),\\
 {\cal S}_1^{(6)} (1,2\,|\,3,4,5,6)&= \frac{b_1}{\Lambda^4 f^4}\Bigg[s_{12} \bigg(s_{13}^2+\left(s_{14}+s_{15}+2 s_{23}+s_{24}+s_{25}+3 s_{34}+s_{35}\right) s_{13}+s_{23}^2\nonumber\\
 &+2 s_{34}^2+s_{14} s_{23}+s_{15} s_{23}+s_{23} s_{24}+s_{23} s_{25}+2 s_{14} s_{34}+s_{15} s_{34}+3 s_{23} s_{34}\nonumber\\
 &+2 s_{24} s_{34}+s_{25} s_{34}+s_{12} \left(s_{13}+s_{23}+2 s_{34}\right)+s_{14} s_{35}+s_{23} s_{35}+s_{24} s_{35}\nonumber\\
 &+2 s_{34} s_{35}\bigg)\Bigg],
 \label{eqapp:6ptdt11}
 \end{align}
 \begin{align}
{\cal S}_1^{(6)} (1,2\,|\,3,4,5,6)&= \frac{b_2}{\Lambda^4 f^4}\Bigg[\frac{1}{2} s_{12}^2 \left(s_{13}+s_{23}\right) +\frac{1}{2} s_{12}\bigg(s_{13}^2+\big(s_{14}+s_{15}+4 s_{23}+3 s_{24}+s_{25}\nonumber\\&+s_{34}+s_{35}\big) s_{13}+s_{23}^2+s_{15} s_{23}+s_{23} s_{24}+s_{23} s_{25}-s_{15} s_{34}+s_{23} s_{34}-s_{25} s_{34}\nonumber\\
&+s_{23} s_{35}+s_{24} s_{35}+s_{14} \left(3 s_{23}+s_{35}\right)\bigg) +s_{14} s_{23}^2+s_{14}^2 s_{23}+s_{14} s_{15} s_{23}\nonumber\\
&+s_{14} s_{23} s_{24}+s_{13}^2 \left(s_{23}+s_{24}\right)+s_{14} s_{23} s_{34}-s_{15} s_{24} s_{34}-s_{14} s_{25} s_{34}\nonumber\\
&-s_{15} s_{25} s_{34}+s_{14} s_{23} s_{35}+s_{14} s_{24} s_{35}+s_{13} \bigg(s_{15} s_{23}+s_{14} \left(2 s_{23}+s_{24}\right)\nonumber\\
&+\left(s_{23}+s_{24}\right) \left(s_{23}+s_{24}+s_{25}+s_{34}+s_{35}\right)\bigg)\Bigg],\\
{\cal S}_1^{(6)} (1,2\,|\,3,4,5,6)&= \frac{b_3}{\Lambda^4 f^4}\Bigg[-s_{12}\bigg(s_{13}^2+s_{13}\left(s_{14}+s_{15}+4 s_{23}+3 s_{24}-s_{25}+s_{34}+s_{35}\right)+s_{23}^2\nonumber\\
&-s_{15} s_{23}+s_{23} s_{24}+s_{23} s_{25}-s_{15} s_{34}+s_{23} s_{34}-s_{25} s_{34}+s_{23} s_{35}+3 s_{24} s_{35}\nonumber\\
&+3 s_{14} \left(s_{23}+s_{35}\right)\bigg) -s_{12}^2 \left(s_{13}+s_{23}\right)-2 \bigg(s_{13}^2\left(s_{23}+s_{24}-s_{25}\right) \nonumber\\
&+s_{13}\big(s_{23}^2+\left(2 s_{24}+s_{34}+s_{35}\right) s_{23}+s_{24}^2-s_{25}^2+s_{14} \left(2 s_{23}+s_{24}\right)-s_{15} s_{25}\nonumber\\&+s_{24} s_{25}+s_{24} s_{34}+2 s_{24} s_{35}-s_{25} s_{35}\big) +s_{14}^2 s_{23}-s_{15} \big(s_{23}^2+s_{15} s_{23}\nonumber\\
&+s_{23}\left(s_{25}+s_{35}\right) +s_{24} s_{34}+s_{25} s_{34}-s_{24} s_{35}\big)+s_{14} \big(s_{23}^2+s_{15} s_{23}\nonumber\\
&+s_{23}\left(s_{24}+s_{34}+2 s_{35}\right) -s_{25} s_{34}+3 s_{24} s_{35}+s_{25} s_{35}\big)\bigg)\Bigg],
 \label{eqapp:6ptdt12}
\end{align}
where $b_0$, $b_1$, $b_2$, and $b_3$ are arbitrary constants. 

\subsubsection{$\bf{\left(1,2,3\,|\,4,5,6\right)}$}
\label{app:6ptdt2}
The double-trace soft blocks ${\cal S}^{(6)} (1,2,3\,|\,4,5,6)$ are invariant under the separate cyclic transformations: $\left(123\right)$, $\left(456\right)$, trace exchange symmetry: $\left(123\right)\leftrightarrow\left(456\right)$  and hermitian conjugation: $(123)\leftrightarrow(321)$ and $(456)\leftrightarrow(654)$. There are three double-trace $6$-pt soft blocks  ${\cal S}^{(6)} (1, 2, 3\,|\,4, 5, 6)$ at ${\cal O}\left(p^6\right)$ that satisfy the above symmetries:
\begin{align}
 {\cal S}_1^{(6)} (1,2,3\,|\,4,5,6)&= \frac{c_0}{\Lambda^4 f^4}\Bigg[s_{12} s_{15} s_{34}+s_{15} s_{23} s_{34}+s_{15} s_{24} s_{34}+s_{15} s_{25} s_{34}+s_{12} s_{24} s_{35}\nonumber\\
 &+s_{15} s_{24} s_{35}+s_{13} \left(s_{14} s_{25}+s_{24} s_{35}\right)+s_{14} \left(s_{23} s_{25}+s_{34} s_{25}+s_{35} s_{25}+s_{24} s_{35}\right)\Bigg],
  \label{eqapp:6ptdt21}
  \end{align}
  \begin{align}
 {\cal S}_1^{(6)} (1,2,3\,|\,4,5,6)&= \frac{c_1}{\Lambda^4 f^4}\Bigg[s_{12} s_{25} s_{34}+s_{14} s_{25} s_{34}+s_{13} \left(s_{15} s_{24}+s_{25} s_{34}\right)+s_{12} s_{14} s_{35}\nonumber\\&+s_{14} s_{23} s_{35}+s_{14} s_{24} s_{35}+s_{14} s_{25} s_{35}+s_{15} \left(s_{23} s_{24}+s_{34} s_{24}+s_{35} s_{24}+s_{25} s_{34}\right)\Bigg],\\
 {\cal S}_1^{(6)} (1,2,3\,|\,4,5,6)&= \frac{c_1}{\Lambda^4 f^4}\Bigg[s_{12}^2\left(s_{13}+s_{23}+s_{34}+s_{35}\right)+s_{12}\bigg(s_{13}^2+s_{13}\big(s_{14}+s_{15}+3 s_{23}\nonumber\\&+2 s_{24}+2 s_{25}+2 s_{34}+2 s_{35}\big) +s_{23}^2+s_{34}^2+s_{35}^2+2 s_{14} s_{23}+2 s_{15} s_{23}+s_{23} s_{24}\nonumber\\&+s_{23} s_{25}+s_{14} s_{34}+2 s_{23} s_{34}+s_{24} s_{34}+s_{15} s_{35}+2 s_{23} s_{35}+s_{25} s_{35}+s_{34} s_{35}\bigg)\nonumber\\&+s_{14} s_{23}^2+s_{15} s_{23}^2+s_{14}^2 s_{23}+s_{15}^2 s_{23}+s_{14} s_{15} s_{23}+s_{14} s_{23} s_{24}+s_{15} s_{23} s_{25}\nonumber\\&+s_{13}^2 \left(s_{23}+s_{24}+s_{25}\right)+s_{14} s_{23} s_{34}-s_{15} s_{24} s_{34}-s_{14} s_{25} s_{34}-s_{15} s_{25} s_{34}\nonumber\\&+s_{15} s_{23} s_{35}-s_{14} s_{24} s_{35}-s_{15} s_{24} s_{35}-s_{14} s_{25} s_{35}+s_{13} \bigg(s_{23}^2+2 s_{14} s_{23}\nonumber\\&+2 s_{15} s_{23}+2 s_{24} s_{23}+2 s_{25} s_{23}+s_{34} s_{23}+s_{35} s_{23}+s_{24}^2+s_{25}^2+s_{14} s_{24}\nonumber\\&+s_{15} s_{25}+s_{24} s_{25}+s_{24} s_{34}+s_{25} s_{35}\bigg)\Bigg],
 \label{eqapp:6ptdt22}
\end{align}
where $c_0$, $c_1$, and $c_2$ are arbitrary constants.
\subsection{Triple-Trace}
\label{app:6pttt}
The triple-trace soft blocks ${\cal S}^{(6)} (1,2\,|\,3,4\,|\,5,6)$ are invariant under the separate cyclic transformations: $\left(12\right)$, $\left(34\right)$, and $\left(56\right)$, trace exchange symmetry: $\left(12\right)\leftrightarrow\left(34\right)$,  $\left(12\right)\leftrightarrow\left(56\right)$, and  $\left(34\right)\leftrightarrow\left(56\right)$,  and hermitian conjugation. There are three triple-trace $6$-pt soft blocks at ${\cal O}\left(p^6\right)$ that satisfy the above symmetries:
\begin{align}
 {\cal S}_1^{(6)} (1,2\,|\,3,4\,|\,5,6)&= \frac{d_0}{\Lambda^4 f^4}\,s_{12} s_{34} s_{56},
 \label{eqapp:6pttt1}
 \end{align}
 \begin{align}
 {\cal S}_1^{(6)} (1,2\,|\,3,4\,|\,5,6)&= \frac{d_1}{\Lambda^4 f^4}\Bigg[s_{34}\bigg(s_{25}\left(s_{13}+s_{14}\right) +s_{12} \left(s_{15}+s_{25}\right)+s_{15} \left(s_{23}+s_{24}+2 s_{25}\right)\bigg) \nonumber\\
&- s_{56}\left(s_{14} s_{23}+s_{13} s_{24}\right) -s_{12} \bigg(s_{13}^2+s_{23}^2+s_{14} s_{23}+s_{15} s_{23}+s_{23} s_{24}+s_{23} s_{25}\nonumber\\
& +s_{15} s_{34}+ s_{12} \left(s_{13}+s_{23}\right)+s_{23} s_{34}+s_{25} s_{34}+ s_{14} s_{35}+s_{23} s_{35}+s_{24} s_{35} \nonumber\\
& +2 s_{34} s_{35}+2 s_{35} s_{46}+s_{13}\left(s_{14}+s_{15}+2 s_{23}+s_{24}+s_{25}+s_{34}+s_{35}\right)\bigg)\Bigg],\\
 {\cal S}_1^{(6)} (1,2\,|\,3,4\,|\,5,6)&= \frac{d_2}{\Lambda^4 f^4}\Bigg[-s_{12}^2\left(s_{13}+s_{23}\right) -s_{12}\bigg(s_{13}^2+s_{23}^2+3 s_{15} s_{23}+s_{23} s_{24}+s_{23} s_{25}\nonumber\\
 &+s_{23} s_{34}+s_{14} \left(s_{23}-s_{35}\right)+s_{23} s_{35}-s_{24} s_{35}\nonumber\\
 &+ s_{13}\left(s_{14}+s_{15}+4 s_{23}+s_{24}+3 s_{25}+s_{34}+s_{35}\right)\bigg) \nonumber\\
&-2 s_{15} s_{23}^2-2 s_{15}^2 s_{23}-2 s_{14} s_{15} s_{23}-2 s_{15} s_{23} s_{25}-2 s_{13}^2 \left(s_{23}+s_{25}\right)\nonumber\\
&-s_{15} s_{23} s_{34}+s_{15} s_{24} s_{34}+s_{14} s_{25} s_{34}-s_{56}\left(s_{14} s_{23}+s_{13} s_{24}\right)-2 s_{15} s_{23} s_{35}\nonumber\\
&+2 s_{14} s_{24} s_{35}+2 s_{15} s_{24} s_{35}+2 s_{14} s_{25} s_{35}-s_{13} \bigg(2 s_{23}^2+2 s_{14} s_{23}+2 s_{24} s_{23}\nonumber\\
&+4 s_{25} s_{23}+2 s_{34} s_{23}
+2 s_{35} s_{23}+2 s_{25}^2+2 s_{24} s_{25}+2 s_{15} \left(2 s_{23}+s_{25}\right)\nonumber\\
&+s_{25} s_{34}+2 s_{25} s_{35}\bigg)\Bigg],
\label{eqapp:6pttt2}
\end{align}
where $d_0$, $d_1$, and $d_2$ are arbitrary constants.

\section{Coding for Computing Soft Blocks}
\label{subsec:computing}
In this Appendix we briefly describe the algorithms used to compute soft blocks in the accompanying {\tt Mathematica} notebook, with the hope to make the notebook more accessible to the readers. 
Up to 8-pt at ${\cal O}(p^{8})$, the soft blocks can be easily computed on a typical laptop computer. For 8-pt and 10-pt at ${\cal O}(p^{10})$, we ran the codes on a computing cluster with 44 CPU's. Using the parallel computing feature in {\tt Mathematica}, it generally takes less than a day  to obtain the soft blocks of a particular trace structure at ${\cal O}(p^{10})$.

The inputs of the algorithm to compute soft blocks are: Adler's zero condition or cyclic invariance, a set of independent monomials, and a set of independent blocks.  The outputs is a set of blocks satisfying the input  condition. Both Adler's zero condition and cyclic invariance can be expressed as transformations of kinematic variables $\{s_{ij} \to \sigma(s_{ij})\}$; see for example Eq.~(\ref{eq:6ptAdler})  for Adler's zero condition. The set of independent monomials are used to extract coefficients a polynomial. At the very start, when no Adler's zero or cyclic conditions have been imposed, the set of independent blocks is the same as the set of independent monomials. 

Denote a set of blocks as $\{\tilde{\cal S}_n(\{s_{ij}\})\}$, $n=1,...,N$, with $N$ the input number of blocks.\footnote{These are {\it not} soft blocks yet, so we put a tilde on ${\cal S}_n$.} The following is the algorithm to obtain soft blocks, schematically,
\begin{enumerate}
	\item Set ${\tilde{\cal S}_n(\{s_{ij}\})}$ equal to the set of independent monomials.
	\item Define a polynomial $p(\{s_{ij}\})=\sum_{n=1}^N c_n \left(\tilde{\cal S}_n(\{\sigma(s_{ij})\}) - \tilde{\cal S}_n(\{s_{ij}\}) \right)$ and extract the coefficients of $p(\{s_{ij}\})$ with respect to the input monomials. Setting the coefficients of the input monomials to zero gives rise to a set of linear equations of the coefficients $\{c_n\}$. 
	\item Solve the linear equations of $\{c_n\}$. The solution is that a subset $D$ of $\{c_n\}$ is expressed as linear functions of the rest of $\{c_n\}$ denoted as $I$.\footnote{$D$ stands for Dependent coefficients and $I$ Independent coefficients} 
	\item Plug the solutions of the previous step back to $\sum_{n=1}^N c_n \tilde{\cal S}_n(\{s_{ij}\})$. Extract the coefficients of $c_n \in I$ and the set of extracted coefficients (polynomials of $\{s_{ij}\}$) is the set of output blocks.
	\item Repeat from step 2 until all Adler's zero conditions and cyclic conditions (also hermitian and anti-hermitian conditions) are imposed. In the end, the set of output blocks is the set of soft blocks.
\end{enumerate}
In Step 2 above, if Adler's zero is to be implemented, then $p(\{s_{ij}\})=\sum_{n=1}^N c_n \tilde{\cal S}_n(\{\sigma(s_{ij})\}) $ where $\sigma$ corresponds to imposing Adler's zero condition in Eq.~(\ref{eq:6ptAdler}). The most time consuming and computational demanding steps are Step 2 and Step 3. Efforts have been made to greatly optimize the notebook so as to obtain the ${\cal O}(p^{10})$ soft blocks. 

\bibliography{references_amp}

\end{document}